\documentclass[5p]{elsarticle}

\usepackage{latexsym,bbold,colordvi}
\usepackage{bm}

\usepackage[fleqn]{amsmath}

\biboptions{sort&compress}

\newcommand{\eqnb}{\begin{equation}}
\newcommand{\eqne}{\end{equation}}

\begin{document}
\begin{frontmatter}

%%%%%%%%%%%%%%%%%%%%%%%%%%%%%%%%%%%%%%%%%%%%%%%%%%%%%%%%%%%%%

\let\I\i
\def\i{\mathrm{i}}
\def\e{\mathrm{e}}
\def\d{\mathrm{d}}
\def\half{{\textstyle{1\over2}}}
\def\thalf{{\textstyle{3\over2}}}
\def\h{{\scriptscriptstyle{1\over2}}}
\def\th{{\scriptscriptstyle{3\over2}}}
\def\fh{{\scriptscriptstyle{5\over2}}}
\def\vec#1{\mbox{\boldmath$#1$}}
\def\svec#1{\mbox{{\scriptsize \boldmath$#1$}}}
\def\oN{\overline{N}}
\def\ttimes{{\scriptstyle \times}}
\def\bm#1{{\pmb{\mbox{${#1}$}}}}

\def\CG#1#2#3#4#5#6{C^{#5#6}_{#1#2#3#4}}
%j1,j2,j3,m1,m2,m3
\def\threej#1#2#3#4#5#6{\left(\begin{array}{ccc}
    #1&#2&#3\\#4&#5&#6\end{array}\right)}

\title{A $U_A(1)$ symmetry restoration scenario supported by the generalized Witten-Veneziano relation and its analytic solution}

\author[pmf]{S. Beni\'c}
\author[pmf]{D. Horvati\' c}
\author[irb]{D. Kekez}
\author[pmf]{D. Klabu\v{c}ar}
\cortext[cor]{Corresponding author.}
\ead{klabucar@phy.hr}
% \preprint{ZTF-EP-07}

\address[pmf]{Physics Department, Faculty of Science,
University of Zagreb, Bijeni\v{c}ka c. 32, 10000 Zagreb, Croatia}
\address[irb]{Rudjer Bo\v{s}kovi\'{c} Institute, P.O.B. 180,
10002 Zagreb, Croatia}
\date{\today}

\begin{abstract}
The Witten-Veneziano relation, or, alternatively, its generalization proposed 
by Shore, facilitates understanding and 
describing the complex of $\eta$ and $\eta^\prime$ mesons. We present 
an analytic, closed-form solution to Shore's equations which gives results 
on the $\eta$-$\eta^\prime$ complex in full agreement 
with results previously obtained numerically.
Although the Witten-Veneziano relation and Shore's equations are related, 
the ways they were previously used in the context of dynamical models to 
calculate $\eta$ and $\eta^\prime$ properties,
were rather different. However, 
with the analytic solution, the calculation can be formulated 
similarly to the approach through the Witten-Veneziano relation,
and with some conceptual improvements. 
In the process, one 
strengthens the arguments in favor of a possible
relation between the $U_A(1)$ and $SU_A(3)$ chiral symmetry 
breaking and restoration. 
To test this scenario, the experiments such as those at RHIC, NICA and
FAIR, which extend the RHIC (and LHC) high-temperature scans also to the
finite-density parts of the QCD phase diagram, should pay particular
attention to the signatures from the $\eta'$-$\eta$ complex indicating
the symmetry restoration.
\end{abstract}


\end{frontmatter}

\section{Introduction}
\label{INTRO}

Among the most intriguing results from RHIC are those on the increased
$\eta'$ multiplicities found in the $\sqrt{s_{N\!N}} = 200$ GeV central
Au+Au reactions \cite{Adler:2004rq,Adams:2004yc}, since Cs\"org\H{o} 
{\it et al.} \cite{Csorgo:2009pa,Vertesi:2009wf,Vargyas:2012ci} established 
that this implied that the 
vacuum value of the $\eta'$ meson mass, $M_{\eta'}=957.8$ MeV, was 
reduced by at least 200 MeV inside the fireball. This was interpreted
as the ``return of the prodigal Goldstone boson" predicted as a signal 
of the $U_A$(1) symmetry restoration \cite{Kapusta:1995ww}. Namely, 
$M_{\eta'}$ is so very high
due to the nonabelian, ``gluon" axial anomaly breaking the $U_A$(1) symmetry 
and so precluding $\eta'$ from being the 9th (almost-)Goldstone boson of QCD.

Nevertheless, it may seem 
somewhat surprising that the $U_A$(1) symmetry restoration
is observed before deconfinement or the restoration of
the [$SU_A(N_f)$ flavor] chiral symmetry of QCD (whose
dynamical breaking results in light, (almost-)Goldstone
pseudoscalar meson octet
$P = \pi^0,\pi^\pm,K^0, {\bar K}^0, K^\pm,\eta$ for $N_f = 3$).
Namely, the $U_A(1)$ symmetry restoration was expected to occur last, at the
temperature ($T$) scale characterizing the pure-gauge, Yang-Mills 
(YM) theory, which is significantly higher (by some 100 MeV) 
than the $T$ scale characterizing the full QCD.
% with quarks. 
For example, consider the Witten-Veneziano relation (WVR) 
\cite{Witten:1979vv,Veneziano:1979ec}, seemingly peculiarly relating 
the four quantities of the full QCD, namely the pion decay constant 
$f_\pi$ and $\eta'$, $\eta$ and $K$-meson masses $M_{\eta',\eta,K}$,
to YM topological susceptibility $\chi_{\rm Y\!M}$:
\begin{equation}
M_{\eta'}^2 + M_\eta^2 - 2 M_K^2 = 2 N_f \,\, \frac{\chi_{\rm Y\!M}}{f_\pi^2} \, .
\label{WittenVenez}
\end{equation}
It shows that the anomalously large mass of $\eta'$, nonzero even in 
the chiral limit due to the breaking of the $U_A(1)$ symmetry, is 
determined by the ratio of 
$\chi_{\rm Y\!M}$ and $f_\pi$.

WVR is well satisfied at $T=0$ for
$\chi_{\rm Y\!M}$ obtained by lattice calculations (e.g.,
\cite{Lucini:2004yh,DelDebbio:2004ns,Alles:2004vi,Durr:2006ky}).
Nevertheless,
the $T$-dependence of $\chi_{\rm Y\!M}$ is such that the 
straightforward extension of Eq. (\ref{WittenVenez}) to $T>0$
\cite{Horvatic:2007qs}, {\it i.e.}, 
replacement of all quantities{\footnote{Throughout this paper,
all quantities are for definiteness assumed at $T=0$ unless their
$T$-dependence is specifically indicated in formulas or in the text.}}
therein by their respective
$T$-dependent versions $M_{\eta'}(T)$, $M_{\eta}(T)$, $M_K(T)$,
$f_\pi(T)$ {\it and} $\chi_{\rm Y\!M}(T)$, then follows the (naive) 
expectation that chiral symmetry restoration occurs significantly
before the $\chi_{\rm Y\!M}(T)$ ``melting" and the partial 
$U_A(1)$ symmetry restoration, which leads to a conflict 
with experiment \cite{Csorgo:2009pa,Vertesi:2009wf}.

Such a conflict is expected
at high $T$, since WVR relates the four full-QCD quantities 
($M_{\eta',\eta,K}$~, $f_\pi$) with $\chi_{\rm Y\!M}$,
% a pure-gauge quantity, {\it i.e.}, 
a quantity from the pure-gauge, YM theory, where one finds a much 
larger resilience to increasing $T$ than in QCD, which contains also 
quark degrees of freedom. We thus conjectured \cite{Benic:2011fv} that 
the experimentally observed $\eta'$ multiplicities can be explained by 
invoking the Leutwyler-Smilga (LS) relation \cite{Leutwyler:1992yt}. 
It expresses the full-QCD topological susceptibility $\chi$ through the 
YM topological susceptibility $\chi_\text{YM}$ (equal  
to $\chi$ in the limit of {\it quenched} QCD), the {\it chiral-limit} 
quark condensate $\langle{\bar q}q\rangle_0$, and 
the current masses $m_q$ of the $N_f=3$ light quark flavors. 
Inverted, and in our notation, the LS relation is
\begin{equation}
\label{LS}
\frac{\chi}{1+\frac{\chi}{\langle{\bar q}q\rangle_0}
{\displaystyle\sum\limits_{q=u,d,s}} \frac{1}{m_{q}} } \, \,
\, \equiv {\widetilde \chi} \, = \, \chi_{\rm Y\!M}
 \qquad ({\rm at } \, \, T=0)\, .
\end{equation}
Ref. \cite{Benic:2011fv} proposed that the presence of 
$\chi_{\rm Y\!M}$ in WVR should be understood in the light of 
the LS relation (\ref{LS}); {\it i.e.},  the successful
zero-temperature WVR is retained, since 
$\chi_{\rm Y\!M} = {\widetilde \chi}$ at $T=0$, 
but at $T > 0$, ${\widetilde \chi}(T)$ should be used instead 
of $\chi_{\rm Y\!M}(T)$, avoiding the mismatch of the
$T$-dependences of QCD and YM theory.
Since ${\widetilde \chi}$ is a combination of the quantities 
of the full QCD, it should be much less $T$-resistant than 
$\chi_{\rm Y\!M}$. Indeed, employing the light-quark-sector
result \cite{Di Vecchia:1980ve,Leutwyler:1992yt} appropriate
for the topological susceptibility{\footnote{Although lattice 
calculations of the topological susceptibility with light quarks 
are much harder than those of $\chi_{\rm Y\!M} = \chi_{\rm quenched}$,
note that recent lattice calculations also yield the full QCD 
topological susceptibility which vanishes in the limit of a vanishing 
quark mass -- for example, see \cite{Bernard:2012fw}.}} of the full QCD,
\begin{equation}
\chi = - \frac{\langle{\bar q}q\rangle_0}
{{\displaystyle\sum\limits_{q=u,d,s}} \frac{1}{m_{q}}} + {\cal C}_m
 \, ,
\label{chi_small_m}
\end{equation}
(where ${\cal C}_m$ denotes corrections of higher orders{\footnote{Nevertheless,
having ${\cal C}_m \neq 0$ is essential so that the LS relation (\ref{LS}) 
with Eq. (\ref{chi_small_m}) can yield relatively large but finite  
$\chi_{\rm Y\!M}$.}} in small $m_q$), 
yields ${\widetilde \chi}(T)$ which falls with $T$ proportionally to the 
chiral quark condensate $\langle{\bar q}q\rangle_0$ \cite{Benic:2011fv}.
This way, the (partial) restoration of $U_A(1)$ symmetry is naturally 
tied to the restoration of the $SU_A(3)$ flavor chiral symmetry and to 
its characteristic temperature $T_{\rm Ch}$.
This scenario enabled Ref. \cite{Benic:2011fv} to provide the 
first explanation of the findings of Cs\"org\H{o} {\it et al.} 
\cite{Csorgo:2009pa,Vertesi:2009wf},
since the anomalous part of the $\eta'$ mass falls together with  
$\langle{\bar q}q\rangle_0(T)$ as $T \to T_{\rm Ch}$.
Some other approaches \cite{Alkofer:2011va,Kwon:2012vb,Lee:2013es} 
have also provided indirect support to this scenario, but it remained 
just a conjecture on the $T$-dependence of WVR, until the present paper. 
The paper is organized as follows. Sec. \ref{AnalyticSolToShoresRelations}
recalls Shore's \cite{Shore:2006mm,Shore:2007yn} generalization of WVR, 
and how it was adapted 
to $q\bar q$ bound-state calculations \cite{Horvatic:2007mi}.
This entails replacing $\chi_{\rm Y\!M}$ by a quantity to which 
the LS relation (\ref{LS}), {\it i.e.}, $\widetilde \chi$, is just a
large-$N_c$ approximation. This confirms the conjecture of 
Ref. \cite{Benic:2011fv}, especially after we present an 
analytic solution of Shore's equations, and discuss its
implications in Sec. \ref{Discussion}. 
We summarize in Sec. \ref{Summary}.

\section{Analytic solution to generalized Witten-Veneziano relations}
\label{AnalyticSolToShoresRelations}

Like the LS relation (\ref{LS}), WVR 
(\ref{WittenVenez}) was derived in the lowest-order approximation 
in the large $N_c$ expansion. Its generalization by Shore is however 
valid to all orders in $1/N_c$ \cite{Shore:2006mm,Shore:2007yn}. 
It consists of several relations,
and the ones pertinent for the present paper are those containing the 
masses of the pseudoscalar nonet mesons:
\begin{eqnarray}
(f^0_{\eta'})^2 M_{\eta'}^2 + (f^0_{\eta})^2 M_\eta^2 =
{1\over3} \bigl(f_\pi^2 M_\pi^2 + 2 f_K^2 M_K^2\bigr) + 6 A~,
\label{eq:bn}
\\
f^0_{\eta'} f^8_{\eta'} M_{\eta'}^2 + f^0_{\eta} f^8_{\eta} M_{\eta}^2 =
{2\sqrt2\over3}\bigl(f_\pi^2 M_\pi^2 - f_K^2 M_K^2\bigr)~,
\label{eq:bo}
\\
(f^8_{\eta'})^2 M_{\eta'}^2 + (f^8_{\eta})^2 M_{\eta}^2 =
-{1\over3}\bigl(f_\pi^2 M_\pi^2 - 4 f_K^2 M_K^2\bigr)~.
\label{eq:bp}
\end{eqnarray}
Here, $A$ is the full QCD topological charge parameter, namely the 
quantity which takes the role of $\chi_\text{YM}$ in WVR,
\begin{equation}
\label{defA}
A \, = \, \frac{\chi}{1+{\chi} (\frac{1}{\langle {\bar u}u \rangle \, m_{u}}
 + \frac{1}{\langle {\bar d}d \rangle \, m_{d}}
+ \frac{1}{\langle {\bar s}s \rangle \, m_{s}})} \, ,
\end{equation}
given by the full QCD topological susceptibility $\chi$, the current quark masses $m_q$, 
and the three condensates $\langle q \bar q \rangle$ which differ from each other for 
different flavors $q=u,d,s$.     Since they are not known well enough, 
Shore himself \cite{Shore:2006mm,Shore:2007yn}, as well as Ref. \cite{Horvatic:2007mi} 
which adapted his generalization 
of WVR to the Dyson-Schwinger (DS) bound-state approach, had to approximate $A$ by 
values of $\chi_\text{YM}$ \cite{Shore:2006mm,Shore:2007yn,Horvatic:2007mi} found 
on lattice. This is a good approximation at least in the large $N_c$ limit, as 
$A = \chi_\text{YM} + {\cal O}({1}/{N_c})$ [consistent with, {\it e.g.}, the 
large-$N_c$ relation (\ref{LS})]. On the other hand, in the chiral limit 
($m_q\to 0, \forall q$), all $\langle q \bar q \rangle$ condensates tend 
to the chiral one, $\langle q \bar q \rangle_0$.
Since this limit is not far from the real world even when the % third, 
strange flavor is included,
%, $q=s$, is included, the 
the possibility of confirming the conjecture of Ref. \cite{Benic:2011fv}, 
and thus of better understanding the experimental results 
\cite{Csorgo:2009pa,Vertesi:2009wf} signaling the restoration of $U_A(1)$ symmetry, 
becomes apparent. Namely, when
$\langle {\bar u}u \rangle, \langle {\bar d}d \rangle, \langle {\bar s}s \rangle
\to \langle q \bar q \rangle_0$, the topological charge parameter $A$ 
(\ref{defA}) reduces to the quantity $\widetilde \chi$ defined by Eq. (\ref{LS}).
This obviously supports our conjecture \cite{Benic:2011fv} that  
$\widetilde \chi(T)$ determines the $T$-dependence for the anomalous mass.

Eqs. (\ref{eq:bn})-(\ref{eq:bp}) take into account that 
$\eta$ and $\eta'$ possess {\it two} decay constants each 
\cite{Gasser:1984gg,Leutwyler98,KaiserLeutwyler98}, {\it i.e.}, $f^0_{\eta},
f^8_{\eta}$  and  $f^0_{\eta'}, f^8_{\eta'}$, since 
$\eta$ and $\eta'$ are mixtures of the $SU(3)$ 
%flavor 
singlet and octet basis states
$\eta_0$ and $\eta_8$. 
These four decay constants 
% $f^0_{\eta}, f^8_{\eta}$, $f^0_{\eta'}$ and $f^8_{\eta'}$,
%%of the physical $\eta$ and $\eta'$ 
can be parametrized in terms of two auxiliary decay constants and 
two angles; {\it e.g.}, the purely octet and singlet decay constants 
$f_8$ and $f_0$, and the mixing angles $\theta_8$ and $\theta_0$:
\begin{equation}
\! \! \left[ \begin{array}{cc}
f^8_\eta &
   f^0_\eta
\\
f^8_{\eta^\prime} &
   f^0_{\eta^\prime}
       \end{array}
\right]
\! = \!
\left[ \begin{array}{cr}
 \cos\theta_8   & - \sin\theta_0
\\
 \sin\theta_8   &   \cos\theta_0
       \end{array}
\right]
\!
\left[ \begin{array}{cc}
\   f_8   &    0   
\\
     0    &   f_0  
       \end{array}
\right] \, .
\label{DEF-f8etaf0eta}
\end{equation}
For realistic quark masses, $\theta_8$ and $\theta_0$ are rather different
both from each other and from $\theta$, the mixing angle of the states 
$\eta_8$ and $\eta_0$ into $\eta$ and $\eta'$
\cite{Gasser:1984gg,Schechter:1992iz,Leutwyler98,KaiserLeutwyler98,FeldmannKrollStech98PRD,FeldmannKrollStech99PLB,Feldmann99IJMPA}. Only in the limit of the 
exact SU(3) flavor symmetry, $\theta_8 = \theta_0 = \theta = 0$.

If, instead of the $SU(3)$ basis states $\eta_0$ and $\eta_8$, one uses the
nonstrange-strange (NS-S) basis, $\eta_\text{NS} = (u\bar u + d\bar d)/\sqrt{2}$
and $\eta_S = s\bar s$, one obtains 
\begin{equation}
 \left[ \begin{array}{cc}
f^{\text{NS}}_\eta &
   f^{\text{S}}_\eta
\\
f^{\text{NS}}_{\eta^\prime} &
   f^{\text{S}}_{\eta^\prime}
       \end{array}
\right]
\! = \!
\left[ \begin{array}{cr}
 \cos\phi_{\text{NS}}   & - \sin\phi_{\text{S}}
\\
 \sin\phi_{\text{NS}}   &   \cos\phi_{\text{S}}
       \end{array}
\right]
\!
\left[ \begin{array}{cc}
\   f_{\text{NS}}   &    0   
\\
     0    &   f_{\text{S}}
       \end{array}
\right]  .
\label{DEF-fNSetafSeta}
\end{equation}
where $f_\text{NS}$ and $f_\text{S}$ are given by the matrix elements of 
$A_\text{NS}$ and $A_\text{S}$, the NS and S axial currents of quarks:
\begin{equation}
\langle 0| A_\text{NS(S)}^{\mu}(x) |\eta_\text{NS(S)}(p)\rangle
=
i f_\text{NS(S)}\, p^\mu e^{-ip\cdot x}~,
\end{equation}
whereas $\langle 0| A_\text{NS}^{\mu}(x) |\eta_\text{S}(p)\rangle = 0 \, ,
\, \langle 0| A_\text{S}^{\mu}(x) |\eta_\text{NS}(p)\rangle = 0$.

Differing just by the choices of bases, these two sets of decay constants 
are simply related ({\it e.g.}, see \cite{Feldmann99IJMPA,Kekez:2000aw}):
\begin{equation}
\left[ \begin{array}{cc}
f^\text{NS}_\eta &
        f^\text{S}_\eta
\\
f^\text{NS}_{\eta^\prime} &
        f^\text{S}_{\eta^\prime}
       \end{array}
\right]
=
\left[ \begin{array}{cc}
f^8_\eta &
        f^0_\eta
\\
f^8_{\eta^\prime} &
        f^0_{\eta^\prime}
       \end{array}
\right]
\left[ \begin{array}{cc}
\frac{1}{\sqrt{3}} & -\sqrt{\frac{2}{3}}
\\
\sqrt{\frac{2}{3}} & \frac{1}{\sqrt{3}}
       \end{array}
\right]~,
\label{TwoMixingAngles:sns-etatap}
\end{equation}
and completely equivalent {\it in principle.} Still, there is a big 
{\it practical difference}: in the $\mathrm{NS}$-$\mathrm{S}$ basis, FKS
\cite{FeldmannKrollStech98PRD,FeldmannKrollStech99PLB,Feldmann99IJMPA} 
managed to recover a scheme with a single angle $\phi$, which also plays the 
familiar role of the state mixing angle describing the rotation of the NS-S 
basis states into the mass (squared) eigenstates - the physical $\eta$ and 
$\eta'$ mesons: 
\begin{equation}
\eta = \cos\phi \, \eta_{\text{NS}}
             - \sin\phi \, \eta_\text{S}~,
\,\,\,\,
\eta^\prime = \sin\phi \, \eta_\text{NS}
             + \cos\phi \, \eta_\text{S}~.
\label{MIXphi}
\end{equation}
Of course,
this is done at the expense of the full generality, but also without
losing essential physics, making reasonable approximations by applying 
the Okubo-Zweig-Iizuka (OZI) rule
\cite{FeldmannKrollStech98PRD,FeldmannKrollStech99PLB,Feldmann99IJMPA};
{\it e.g.}, $f_\text{NS} f_\text{S} \sin(\phi_\text{NS}-\phi_\text{S})$
differs from zero just by an OZI-suppressed term \cite{Feldmann99IJMPA}.
Neglecting it therefore implies $\phi_\text{NS} = \phi_\text{S}$.
That is, applications of the OZI rule lead to the FKS approximation
scheme \cite{FeldmannKrollStech98PRD,FeldmannKrollStech99PLB,Feldmann99IJMPA},
which exploits the practical difference between the parameterizations  
(\ref{DEF-f8etaf0eta}) and (\ref{DEF-fNSetafSeta}):  $\theta_8$ and 
$\theta_0$ much differ from each other and from the $\eta_8$-$\eta_0$ 
{\it state} mixing angle $\theta \approx (\theta_8 + \theta_0)/2$, but
the NS-S decay-constant mixing angles are very close 
to each other and both can be approximated by the state mixing angle:
$\phi_\text{NS} \approx \phi_\text{S} \approx \phi$. It is thus a reasonable 
approximation to use only this one angle, $\phi$, and express 
(see, {\it e.g.}, \cite{Feldmann99IJMPA,Kekez:2000aw,Horvatic:2007mi})
the physical $\eta$-$\eta'$ decay constants as
\begin{equation}
\! \! \left[ \begin{array}{cc}
f^8_\eta &
   f^0_\eta
\\
f^8_{\eta^\prime} &
   f^0_{\eta^\prime}
       \end{array}
\right]
\! = \!
\left[ \begin{array}{cr}
f_\text{NS} \cos\phi & -f_\text{S} \sin\phi
\\
f_\text{NS} \sin\phi &  f_\text{S} \cos\phi
       \end{array}
\right]
\!
\left[ \begin{array}{cc}
\frac{1}{\sqrt{3}} & \sqrt{\frac{2}{3}}
\\
-\sqrt{\frac{2}{3}} & \frac{1}{\sqrt{3}}
       \end{array}
\right].
\label{matrixEqLast}
\end{equation}
This result of the FKS scheme was inserted in Shore's equations 
(\ref{eq:bn})-(\ref{eq:bp}) already in Ref. \cite{Horvatic:2007mi}, 
but only numerical solutions were found there (after some additional 
assumptions, see below).  In contrast, now we  
have found analytic, closed-form solutions of combined Eqs. 
(\ref{eq:bn})-(\ref{eq:bp}) and (\ref{matrixEqLast}) for $\phi$ and the 
masses of $\eta$ and $\eta'$.  
The relevant set of solutions, where $M_{\eta^\prime} > M_{\eta}$, is:
\begin{eqnarray}
\label{1phiSolqqbar}
\tan\phi =
\frac{- f_\text{NS}^2 + 2 f_\text{S}^2}{2\sqrt{2}\, f_\text{NS} \, f_\text{S}} \, -
\, \qquad \qquad \qquad \qquad \qquad \quad  \quad
\\ 
\frac{2 f_\text{NS}^2 f_K^2 M_K^2 - f_\text{NS}^2 f_\pi^2 M_\pi^2
- f_\text{S}^2 f_\pi^2 M_\pi^2 - \Delta}{4\sqrt{2}\, A \, f_\text{NS} \, f_\text{S} } \, ,
\nonumber 
%\label{1phiSolqqbar}
\end{eqnarray}
\begin{eqnarray}
M_{\eta^\prime (\eta)}^2 \, = \, \frac{A}{f_\text{S}^2} \, 
+ \, \frac{2A}{f_\text{NS}^2} \,  +
\qquad \qquad\qquad \qquad \qquad \quad \quad 
\\  \frac{ 2 f_\text{NS}^2 f_K^2 M_K^2 - f_\text{NS}^2 f_\pi^2 M_\pi^2
+ f_\text{S}^2 f_\pi^2 M_\pi^2 + (-) \Delta}{2 f_\text{NS}^2 f_\text{S}^2 } \, ,
\nonumber
\label{1qqbarMetasSols}
\end{eqnarray}
\begin{eqnarray}
\Delta^2 \,  = \, 32 \, A^2 \, f_\text{NS}^2 \, f_\text{S}^2 \,  + 
\, \qquad \qquad \qquad \qquad \qquad \quad \quad \quad  
\\
\quad  \left[ 2A(f_\text{NS}^2  - 2 f_\text{S}^2) + 2 f_K^2 f_\text{NS}^2 M_K^2 - 
f_\pi^2 (f_\text{NS}^2+f_\text{S}^2) M_\pi^2  \right]^2 \, . \nonumber
\label{1qqbarDelta}
\end{eqnarray}
The major obstacle to evaluating these results may seem to be the 
lack of information{\footnote{Recently, Ref. \cite{Michael:2013vba} 
presented a lattice analysis of $f_\text{NS}$ and $f_\text{S}$ (denoted by $f_{l}$ 
and $f_{s}$ there), which is nevertheless still affected by residual
lattice artefacts, quark mass dependence and chiral perturbation
theory approximation used there.}} 
on $f_\text{NS}$ and $f_\text{S}$, the decay constants of 
the unphysical pseudoscalars $\eta_\text{NS}$ and $\eta_\text{S}$. However,
the guidance is provided by the nature of the FKS scheme, which
neglects OZI-violating contributions, {\it i.e.}, possible 
gluonium admixtures in $\eta_\text{NS}$ and $\eta_\text{S}$. It is then
reasonable to treat them as pure $q\bar q$ states, whereby 
$f_\text{NS}=f_{u\bar u}=f_{d\bar d}=f_\pi$ (in the isospin symmetry limit), 
and $f_\text{S}=f_{s\bar s}$,
the decay constant of the fictitious $s\bar s$ pseudoscalar meson. 
Then the analytic, closed-form solutions (\ref{1phiSolqqbar}), 
(\ref{1qqbarMetasSols}) and (\ref{1qqbarDelta}) become
\begin{equation}
\tan\phi =
\frac{- f_\pi^2 + 2 f_{s\bar s}^2}{2\sqrt{2}\,f_\pi \, f_{s\bar s}}  - 
\frac{2 f_K^2 f_\pi^2 M_K^2 - f_\pi^4 M_\pi^2
- f_\pi^2 f_{s\bar s}^2 M_\pi^2 - \Delta}{4\sqrt{2}\, A \, f_\pi \, f_{s\bar s} } ,
\label{phiSolqqbar}
\end{equation}
\begin{eqnarray}
\label{qqbarMetasSols}
M_{\eta^\prime (\eta)}^2 \, = \, \frac{A}{f_{s\bar s}^2} \, 
+ \, \frac{2A}{f_\pi^2} \,  +
\qquad \qquad\qquad \qquad \qquad \quad \quad 
\\  \frac{ 2 f_K^2 f_\pi^2 M_K^2 - f_\pi^4 M_\pi^2
+f_\pi^2 f_{s\bar s}^2 M_\pi^2 + (-) \Delta}{2 f_\pi^2 f_{s\bar s}^2 }
 \, , 
\nonumber
\end{eqnarray}
\vskip -8mm  
\begin{eqnarray}
\Delta^2 \,  = \, 32 \, A^2 \, f_\pi^2 \, f_{s\bar s}^2 \,  + 
\, \qquad \qquad \qquad \qquad \qquad \qquad \quad \quad  
\\
\quad \left\{ - 2A(f_\pi^2  - 2f_{s\bar s}^2)  + f_\pi^2 [ -2 f_K^2 M_K^2 + 
(f_\pi^2+f_{s\bar s}^2) M_\pi^2 ] \right\}^2 . \nonumber
\label{qqbarDelta}
\end{eqnarray}
The situation %in agreement with the OZI rule, 
that $\eta_\text{NS}$ and 
$\eta_\text{S}$ are pure $q\bar q$ states, so that $f_\text{NS}$=$f_\pi$ 
and $f_\text{S}$=$f_{s\bar s}$, is realized, for example,
in the DS approach in the rainbow-ladder approximation (RLA).
There, mesons are pure 
$q\bar q$ solutions (of Bethe-Salpeter equations), without any 
gluonium admixtures, which would be prominent possible sources of 
OZI violations. The FKS scheme is well-suited for the usage in
such a context which is in agreement with the OZI rule.
%The FKS scheme is well-suited for the usage in the DS approach 
%in the rainbow-ladder approximation (RLA), where mesons are pure 
%$q\bar q$ solutions (of Bethe-Salpeter equations), without any 
%gluonium admixtures, which are prominent possible sources of 
%OZI violations.
In a bound-state approach, notably the DS approach used in 
Ref. \cite{Horvatic:2007mi}, the decay constants are quantities 
calculated from the $q\bar q$ substructure of mesons. 
Ref. \cite{Horvatic:2007mi} used three different dynamical models 
for the nonperturbative gluon interactions  (in the DS gap and 
Bethe-Salpeter equations) yielding $q\bar q$ meson solutions 
reproducing well the empirical light meson masses and decay constants
including the presently important $M_\pi$, $M_K$, $f_\pi$ and $f_K$.
Along with the auxiliary but unphysical $f_{s\bar s}$, these results
enabled the numerical solutions describing well the $\eta$-$\eta'$
complex in Ref. \cite{Horvatic:2007mi}. Now, the same model results 
used in the analytic solutions (\ref{phiSolqqbar})-(\ref{qqbarMetasSols}) 
reproduce accurately these numerical solutions of Ref. \cite{Horvatic:2007mi}
for all DS models used, and for the same lattice values for $\chi_\text{YM}$ 
used in  Ref. \cite{Horvatic:2007mi} in the approximation 
$A\approx \chi_\text{YM}$. 
(For comparisons with Ref. \cite{Horvatic:2007mi}, we use the same 
$\chi_\text{YM}$ to evaluate Table I, {\it i.e.}, 
{\it the weighted average} $\chi_\text{YM} = (0.1757 \, \rm GeV)^4$ as in Refs.
\cite{Kekez:2005ie,Horvatic:2007qs} and $\chi_\text{YM} = (0.191 \, \rm GeV)^4$ 
\cite{DelDebbio:2004ns} (used by Shore \cite{Shore:2006mm,Shore:2007yn}.)

Moreover, the need for a specific dynamical model to evaluate the 
auxiliary, unphysical quantity $f_{s\bar s}$ can be circumvented,
%by noting that 
since the chiral expansion indicates that it is a good 
approximation to express $f_{s\bar s} = 2 f_K - f_\pi$. Then the
analytic solutions (\ref{phiSolqqbar})-(\ref{qqbarMetasSols}) can be
evaluated by inserting exclusively the empirical values of the masses 
$M_\pi, M_K$ and decay constants $f_\pi, f_K$, yielding a
model-independent description of the $\eta$-$\eta'$ complex
and suffering theoretical uncertainties only due to choosing
$A\approx\chi_\text{YM}$ and the FKS scheme, including the OZI rule. 

The results obtained in this way are summarized in 
Table \ref{TABLE}, showing they are quite similar to those 
%the results 
of Ref. \cite{Horvatic:2007mi} for all rather different 
bound-state models used there.

\begin{table}[t]
\begin{center}
\begin{tabular}{|c|c|}
\hline
$M_{\pi^0}$     & 134.977 \\
$M_K$           & 493.677 \\
$f_{\pi^-}$     & 92.2138 \\
$f_K$           & 110.379 \\
% $A$             & 175.7$^4$ \\
% $A$             & 191$^4$ \\
\hline
\end{tabular}
\hspace*{10mm}
%\begin{center}
\begin{tabular}{|c||c|c|}
\hline
 $A^{1/4}$              &   175.7 	& 191  \\
\hline
 $M_\eta$               &   488.4 	& 503.2 \\
 $M_{\eta^\prime}$      &   832.7	& 949.8 \\
 $\phi$                 & 45.07$^\circ$	& 50.92$^\circ$ \\
 $\theta$               &-9.664$^\circ$ &-3.816$^\circ$ \\
%\hline
 $\theta_0$             &-0.341$^\circ$	& 5.507$^\circ$ \\
 $\theta_8$             &-18.03$^\circ$	&-12.18$^\circ$ \\
 $f_0$                  & 105.7  	& 105.7  \\
 $f_8$                  & 117.7 	& 117.7  \\
 $f_\eta^0$             & 0.630 	& 10.15  \\
 $f_{\eta^\prime}^0$    & 105.7 	& 105.2  \\
 $f_\eta^8$             & 111.9 	& 115.0  \\
 $f_{\eta^\prime}^8$    & -36.43 	& -24.84 \\
\hline
\end{tabular}
\end{center}
\caption{The values adopted for $A$ are the two lattice values of 
$\chi_\text{YM}$ used in Ref.~\cite{Horvatic:2007mi}. The other inputs
(in the small table on the left, taken from Ref.~\cite{PDG2011}) 
are the experimental values for $M_{\pi^0,K}$, $f_{\pi^-,K}$.
% $M_K$, $f_{\pi^-}$ and $f_K$. 
Everything else, starting with $M_\eta$ and $M_{\eta^\prime}$, 
are the calculated quantities. All quantities are given in MeV, except 
the angles $\phi, \theta, \theta_0$ and $\theta_8$, which are in degrees.
  }
\label{TABLE}
\end{table}

\section{Discussion of results}
\label{Discussion}

\noindent
For the choice of larger $\chi_\text{YM}$, the results in Table \ref{TABLE}
are also reasonably close to the $\eta$-$\eta'$ studies, such as 
\cite{Kekez:2000aw,Kekez:2005ie,Klabucar:1997zi}, using the same 
dynamical DS models as Ref. \cite{Horvatic:2007mi} to get the pion and 
kaon masses and decay constants, but the standard $\eta$-$\eta'$ mass 
matrix in conjunction with WVR to describe $\eta$-$\eta'$ complex.

Thanks to the existence of the analytic solutions
(\ref{1phiSolqqbar})-(\ref{qqbarMetasSols}),
we can now understand both the similarities and differences between 
the two $\eta$-$\eta'$ descriptions: the one using ({\it e.g.,} in
\cite{Kekez:2005ie,Benic:2011fv}) the standard WVR, and the other, using 
Shore's generalization \cite{Shore:2006mm,Shore:2007yn} thereof
in conjunction with the FKS scheme, as in Ref. \cite{Horvatic:2007mi}
and here. 

In the latter approach, the $\eta$-$\eta'$ mass matrix was not needed 
\cite{Horvatic:2007mi}, but it can readily be constructed; its matrix
elements in the NS-S basis are:
\begin{eqnarray}
\label{NS}
\!\!\!\!\!\!\!\!\!\!\!\!\!\!  
M_{\text{NS}}^2 &=& M_\eta^2 \cos^2\!\phi + M_{\eta^\prime}^2 \sin^2\!\phi 
= M_\pi^2 + \frac{4A}{f_\pi^2}  \quad \quad  \qquad
\\
M_{\text{S}}^2 &= & M_\eta^2 \sin^2\!\phi + M_{\eta^\prime}^2 \cos^2\!\phi 
\nonumber
\\
\label{S}   &=& \frac{1}{f_{s\bar{s}}^2} 
  [ 2 f_K^2 M_K^2 - f_\pi^2 M_\pi^2 ] + \frac{2A}{f_{s\bar{s}}^2}
\\
\label{NSS}
M_{\text{NSS}}^2 &=& \sin\phi \, \cos\phi \, (M_\eta^2 - M_{\eta^\prime}^2)
= \frac{2\sqrt{2}A}{f_\pi f_{s\bar{s}}}
\end{eqnarray}
where the second equalities in the Eqs. (\ref{NS}), (\ref{S}) and
(\ref{NSS}) are obtained through inserting the analytic 
solutions (\ref{1phiSolqqbar}), (\ref{1qqbarMetasSols}) and 
(\ref{1qqbarDelta}) with $f_\text{NS} = f_\pi$ and $f_\text{S} = f_{s\bar s}$.
Using the DGMOR relations like Shore for $f_\pi^2 M_\pi^2$ and
$f_K^2 M_K^2$, enables one to express the decay constant and mass
of the unphysical $s\bar s$ almost-Goldstone pseudoscalar as
\begin{equation}
\label{f2M2piKssbarConn}
2 f_K^2 \, M_K^2 - f_\pi^2 \, M_\pi^2 = f_{s\bar{s}}^2 \, M_{s\bar{s}}^2 \, ,
\end{equation}
whereby Eq. (\ref{S}) becomes 
$\, \, M_{\text{S}}^2 = M_{s\bar{s}}^2 + 2 A /f_{s\bar{s}}^2$.

The $\eta$-$\eta'$ mass matrix is \cite{Kekez:2000aw,Kekez:2005ie}
\begin{equation}
\label{massMatrix}
{\hat M}^2 = \left[
        \begin{array}{cc} M_{\text{NS}}^2 & M_{\text{NSS}}^2 \\
                        M_{\text{NSS}}^2 & M_{\text{S}}^2 
        \end{array}
      \right]
    = \left[
        \begin{array}{cc} M_\pi^2+2\beta & \sqrt{2}\beta X \\
                        \sqrt{2}\beta X  & M_{s\bar{s}}^2+\beta X^2 
        \end{array}
      \right]  \, ,
\end{equation}
where $X$ is the flavor $SU(3)$-breaking parameter, and 
$\beta \equiv \Delta M_{\eta_0}/3$ denotes $\frac{1}{3}$ of the 
$U_A(1)$-anomalous, chiral-limit-nonvanishing part of the mass of the 
flavor singlet pseudoscalar $\eta_0$ in the  $SU(3)$-symmetric limit 
($X=1$). 

The matrix (\ref{massMatrix}) implies the $\eta$ and $\eta'$ masses
\begin{eqnarray}
M_{\eta^\prime (\eta)}^2 &=& \frac{M_{\text{NS}}^2+M_{\text{S}}^2 +(-) 
\sqrt{(M_{\text{NS}}^2-M_{\text{S}}^2)^2+8\beta^2 X^2}}{2} \, ,
%(M_{\text{NS}}^2+M_{\text{S}}^2)
%-\frac{1}{2}\sqrt{(M_{\text{NS}}^2-M_{\text{S}}^2)^2+8\beta^2 X^2}
\nonumber
\end{eqnarray}
which can also be obtained from the closed-form solutions 
(\ref{1phiSolqqbar}), (\ref{1qqbarMetasSols}) and (\ref{1qqbarDelta}) 
using Eqs. (\ref{NS})-(\ref{f2M2piKssbarConn}), providing a good
consistency check.

One of the advantages of the present approach is that the comparison of 
the matrix elements (\ref{NS}), (\ref{S}) [inserting (\ref{f2M2piKssbarConn})] 
and (\ref{NSS})  with the matrix (\ref{massMatrix}) shows that 
\begin{equation}
X=\frac{f_\pi}{f_{s\bar{s}}} ~,\,\,\,\,  \quad
 \beta = \frac{2A}{f_\pi^2} \equiv \beta_{\text{S}}
\label{Xbeta}
\end{equation}
follows necessarily. In contrast, Eq. (\ref{Xbeta}) for $X$ is usually 
just an educated estimate \cite{lastFootnote}.
\cite{Feldmann99IJMPA,FeldmannKrollStech98PRD,Kekez:2000aw,Kekez:2005ie}. 
Eq. (\ref{Xbeta}) for $\beta$ also explains why one needs higher values 
of $\chi_\text{YM}$ (if one approximates $A\approx \chi_\text{YM}$) than in the 
approach employing WVR \cite{Kekez:2005ie,Benic:2011fv}.
Namely, there the mass matrix yields $\beta$
which is larger for the same $\chi_\text{YM}$. That is,
\begin{equation}
\beta_{\text{WV}} \, = \, \frac{6 \, \chi_\text{YM}}{(2+X^2) \, f_\pi^2} 
\, > \, \frac{2 \, \chi_\text{YM}}{f_\pi^2} \, ,
\end{equation}
since $X < 1$ for any realistic flavor symmetry breaking.

\section{Summary and outlook}
\label{Summary}

\noindent
Shore's generalization \cite{Shore:2006mm,Shore:2007yn} of WVR provides 
a description of the $\eta$-$\eta'$ complex which, in its original form, 
is valid to all orders in the large-$N_c$ expansion.
We have presented the analytic, closed-form solutions
%(\ref{1phiSolqqbar})-(\ref{qqbarMetasSols}) 
of Shore's equations (\ref{eq:bn})-(\ref{eq:bp}) combined with the FKS 
scheme. This was previously solved only numerically \cite{Horvatic:2007mi} 
%(in the context of several $q\bar q$ bound-state models). 
(for several $q\bar q$ bound-state models), while now we have 
closed-form, analytic expressions (\ref{1phiSolqqbar})-(\ref{qqbarDelta})
%for $M_\eta$, $_{\eta'}$, and $\phi$  
for the masses and the mixing angle in the $\eta$-${\eta'}$ complex,
leading to the mass matrix elements (\ref{NS})-(\ref{NSS}). They show explicitly, 
{\it e.g.}, why the flavor breaking is necessarily given by $X=f_\pi/f_{s\bar s}$
and how the full QCD topological charge parameter $A$ (\ref{defA}) replaces 
the YM topological susceptibility $\chi_\text{YM}$ appearing in the standard WVR.
In general, both the present $\eta$-$\eta'$ description and the corresponding 
numerical ones in Ref. \cite{Horvatic:2007mi}, are much better understood now, 
as the analytic solutions have in the previous section exposed clearly 
both similarities and differences with respect to the descriptions 
of the $\eta$-$\eta'$ complex through the mass matrix and standard WVR (in, 
{\it e.g.}, \cite{Kekez:2005ie,Benic:2011fv}).   Obviously,
some of the generality of the original Shore's approach has been reduced
due to the approximations present in the FKS scheme. This scheme is however 
well founded and, through numerous phenomenological applications, has also 
been proven to preserve the essential physics of the $\eta$-$\eta'$ complex, 
which in this context is also an argument for reliability of the $\eta$-$\eta'$ 
description exploiting WVR \cite{Kekez:2005ie,Benic:2011fv}.

%What is more important about the similarities
%%demonstration of the almost-identity 
%of these two descriptions is that 
It is important to note that in the present paper, the YM topological 
susceptibility $\chi_\text{YM}$ is used (at $T=0=\mu$) only as an 
%the large-$N_c$ 
approximation of the full QCD topological charge parameter $A$ (\ref{defA}). The 
latter is, however, not a pure-gauge quantity, but a full QCD quantity, and the 
LS (\ref{LS}) quantity $\widetilde \chi$ is its 
%large-$N_c$ 
approximation recovered from $A$ (\ref{defA}) by replacing $\langle u\bar u\rangle, 
\langle d\bar d\rangle,\langle s\bar s\rangle \to  \langle q\bar q\rangle_0$. 

This relationship between $A$, Eq. (\ref{defA}), appearing as the fundamental 
quantity of the WVR generalization \cite{Shore:2006mm,Shore:2007yn}, and 
$\widetilde \chi$, Eq. (\ref{LS}), supports our explanation \cite{Benic:2011fv} of 
the data on the enhanced $\eta'$-multiplicity \cite{Csorgo:2009pa,Vertesi:2009wf} in 
RHIC experiments at $T>0$, where we replace the $T$-dependence of $\chi_{\rm Y\!M}$ 
by that of $\widetilde \chi(T)$ (\ref{LS}), which is, in essence, the $T$-dependence
of the chiral quark condensate $\langle q\bar q\rangle_0(T)$. 
% Most importantly, 
Such relationship of $U_A(1)$ symmetry breaking to the order parameter of dynamical 
chiral symmetry breaking indicates more strongly the possibility that similar 
experimental signals of $U_A(1)$ symmetry restoration be observed in experiments at 
finite matter density ($\mu > 0$) \cite{NICAwpJan2014}. Namely, the quark condensate 
$\langle q\bar q (T,\mu)\rangle$ should drop significantly not only with $T$, but 
also after the chemical potential $\mu$ exceeds some critical value. This motivates 
the theoretical work \cite{inPreparation} on extending the approach of Ref. 
\cite{Benic:2011fv} to $\mu>0$. 

Previous experimental studies at RHIC have already been extended from the 
high-temperature regime also to the finite density ({\it e.g.,} see 
\cite{Kumar:2012np}), and more studies, including detailed scans of the $\mu$-$T$ 
QCD phase diagram, are planned at RHIC, GSI/FAIR, and NICA \cite{NICAwpJan2014}. 
The present paper stresses it is important that such experiments look for 
signatures (primarily related to the $\eta'$-$\eta$ complex) that would 
test the scenario of Ref. \cite{Benic:2011fv} (and related ideas 
\cite{Kwon:2012vb,Lee:2013es,Jiang:2012wm,Mitter:2013fxa,Sakai:2013nba}) 
on the relationship between the $U_A(1)$ and $SU_A(3)$ chiral symmetry
breaking and restoration.

{\bf Acknowledgment:} 
The authors acknowledge the partial support of 
COST Action MP1304  NewCompStar.

\bibliographystyle{elsarticle-num}

\end{document}